\documentclass[sigconf]{acmart}
\acmConference[MSR 2026]{MSR '26: Proceedings of the 23rd International Conference on Mining Software Repositories}{April 2026}{Rio de Janeiro, Brazil}
\usepackage{enumitem}
\usepackage{url}
\usepackage{hyperref}
\usepackage{multirow}
\usepackage{enumitem}

\copyrightyear{2026}
\acmYear{2026}
\setcopyright{cc}
\setcctype{by-nc-nd}
\acmConference[MSR '26]{23rd International Conference on Mining Software Repositories}{April 13--14, 2026}{Rio de Janeiro, Brazil}
\acmBooktitle{23rd International Conference on Mining Software Repositories (MSR '26), April 13--14, 2026, Rio de Janeiro, Brazil}
\acmPrice{}
\acmDOI{10.1145/3793302.3793578}
\acmISBN{979-8-4007-2474-9/2026/04}

\AtBeginDocument{%
  }

\begin{document}

\title{Mining Type Constructs Using Patterns in AI-Generated Code}

\author{Imgyeong Lee}
\authornote{Both authors contributed equally to this research.}
\email{imgyeong@ualberta.ca}
\affiliation{%
  \institution{University of Alberta}
  \city{Edmonton}
  \state{Alberta}
  \country{Canada}
}

\author{Tayyib Ul Hassan}
\authornotemark[1]
\email{tayyibul@ualberta.ca}
\affiliation{%
  \institution{University of Alberta}
  \city{Edmonton}
  \state{Alberta}
  \country{Canada}
}

\author{Abram Hindle}
\email{abram.hindle@ualberta.ca}
\affiliation{%
  \institution{University of Alberta}
  \city{Edmonton}
  \state{Alberta}
  \country{Canada}
}
\begin{abstract}
Artificial Intelligence (AI) increasingly automates various parts of the software development tasks. Although AI has enhanced the productivity of development tasks, it remains unstudied whether AI essentially outperforms humans in type-related programming tasks, such as employing type constructs properly for type safety, during its tasks. Moreover, there is no systematic study that evaluates whether AI agents overuse or misuse the type constructs under the complicated type systems to the same extent as humans. In this study, we present the first empirical analysis to answer these questions in the domain of TypeScript projects. Our findings show that, in contrast to humans, AI agents are \textit{9×} more prone to use the `\texttt{any}' keyword. In addition, we observed that AI agents use advanced type constructs, including those that ignore type checks, more often compared to humans. Surprisingly, even with all these issues, Agentic pull requests (PRs) have \textit{1.8×} higher acceptance rates compared to humans for TypeScript. We encourage software developers to carefully confirm the type safety of their codebases whenever they coordinate with AI agents in the development process.
\end{abstract}

\ccsdesc[300]{Software and its engineering~Software verification and validation}
\ccsdesc[300]{Software and its engineering~Software maintenance tools}
\ccsdesc[100]{Software and its engineering~Programming teams}
\ccsdesc[300]{Computing methodologies~Artificial intelligence}
\keywords{TypeScript, Type Safety, AI Agents, Large Language Models (LLMs), Empirical Study, Software Quality, Empirical Software Engineering}

\maketitle

\section{Introduction}
TypeScript is one of the most widely used programming languages today. According to the survey with 31,771 participants, 43.6\% of all respondents and 48.8\% of professional developers reported using TypeScript as one of their primary languages~\cite{stackoverflow2025}. One main reason for its popularity is efficiency. TypeScript is a superset of JavaScript that introduces a static type system designed to minimize JavaScript’s major shortcomings like runtime errors~\cite{jansen2016TypeScript}. This characteristic has demonstrated significant improvements over JavaScript in terms of task completion time and error identification, which can be attributed in part to the benefits of its static type system~\cite{fischer2015empirical}. Although TypeScript offers efficiency, the resolution of type errors remains a major challenge in TypeScript development due to its complex type system~\cite {wang2025empirical}.

Concurrently, Artificial Intelligence (AI) agents for programming have rapidly improved productivity in software development. Recent studies report that the use of AI agents can accelerate the implementation of requirements by up to 65\% by utilizing prompt-based interaction and automated code generation~\cite{weber2024significant}. Beyond automatic code generation, AI agents have also demonstrated strong performance in creating PRs for development tasks such as debugging, refactoring, and documentation~\cite {li2025aidev}. As a result, the adoption of AI-assisted programming, including the usage of multi-agent Large Language Models (LLMs), has become a noticeable trend among software developers~\cite{manish2024autonomous, he2025llm}. 

As the use of AI agents in software development becomes increasingly prevalent, researchers have begun to analyze the quality of code generated by these agents. Prior studies have shown that AI-generated code can introduce critical bugs, increase the prevalence of code smells, and ultimately degrade software quality~\cite{siddiq2024quality}. However, to the best of our knowledge, the usage of type constructs, including risky assertions, in AI-generated code has not yet been systematically analyzed.

This study investigates how AI agents exhibit type constructs (hereafter also referred to as type-related features) in TypeScript projects. We not only examine the presence of such features but also analyze how frequently those are introduced, and whether they are overused in ways that can degrade code quality. For example, if AI agents use a method to bypass compile-time checks carelessly, they can introduce subtle runtime errors, which may cause difficulties for maintainability in the future. We compare the type-related features exhibited by AI agents in their code to those exhibited by humans to observe the differences. In this paper, we define a ``type-related pattern'' as the type-safety mechanisms that do not include escape hatches such as \texttt{any} type keyword or non-null assertion(!)~\cite{bogner2022type}. ``Type-related anti-pattern'' refers to the pattern that can lead to a runtime error by bypassing type checks during compile time~\cite{vanderkam2024effective}. We investigate these because they can negatively impact software maintainability.

The motivating research question that we want to answer through this paper is: ``How do AI coding agents differ from human developers in their use of type systems and their impact on code quality?'' We hypothesized that AI agents would exploit type systems to bypass the tests for the fast software development process, instead of solving the actual problems in the codebases. To confirm their behavior, we formulated this main question into three sub-questions:

\begin{enumerate}[label=\textbf{RQ\arabic*}:, leftmargin=*, align=left]
    \item Do AI agents actually introduce type-related issues? If so, how frequently do they introduce them?
    \item How do AI agents and human developers differ in their use of type-related features?
    \item How do Agentic PRs compare to human developers in acceptance rate?
\end{enumerate}

\section{Background}
Static type systems catch type errors early because they execute type checks at compile time. The explicit type annotations also improve code maintainability~\cite{prechelt2002controlled}. However, they can be less flexible in situations where certain values can only be typed at runtime~\cite{baars2002typing}. In these cases, dynamic type systems have a clear advantage. Moreover, dynamic type systems allow developers to be less constrained by type considerations during the early design phase, which often leads to significantly faster prototyping compared to static type systems~\cite{hanenberg2010experiment, stuchlik2011static}.
Due to these characteristics, some programming languages have attempted to unify the two type systems to leverage the benefits of both~\cite{garcia2016design}. Notably among widely used programming languages, TypeScript is a representative example of an attempt to incorporate both type systems into a single programming language. However, TypeScript still involves design trade-offs that break the transitivity of assignment compatibility, potentially leading to degraded code quality when its type constructs are used carelessly~\cite{bierman2014}.

Beyond the theoretical considerations, empirical evidence also suggests that excessive use of the \texttt{any} type has practical consequences. Bogner and Merke analyzed the correlation between the frequency of \texttt{any} type usage and the code quality~\cite{bogner2022type}. Surprisingly, their findings indicate that \texttt{any} type keyword has negative implications, compromising not only soundness but also code quality. They found that the projects with lower frequencies of the \texttt{any} keyword exhibited measurably higher code quality. Although avoiding the \texttt{any} keyword may not guarantee bug-free code, avoiding the \texttt{any} type still improves the maintainability and readability of the code~\cite{bogner2022type}.

While TypeScript provides advanced type-related features to bridge the gap between static and dynamic typing, navigating this trade-off involves balancing compiler strictness with developer convenience, which is a task that requires a deep understanding of the codebases. As AI agents are increasingly integrated into development workflows, investigating their behaviors in real-world projects is essential to understanding the technical debt risks associated with AI-driven development.

\section{Methodology}
All code, prompts, and filtered datasets used in this study are available on Zenodo~\cite{lee_2026_19685727}.

\subsection{Dataset Collection}
We use the AIDev dataset (commit \texttt{eee0408}, accessed on 2025-10-26) for this study~\cite{li2025aidev}.

\subsection{Dataset Filtering Framework}
Our initial dataset contained general PRs that may or may not involve type-related changes. Since our research focuses on type system usage patterns, we needed to extract only PRs with type-related modifications. We designed a hierarchical two-stage filtering pipeline: (1) rule-based regex parsers for efficient initial filtering, and (2) a multi-agent LLM system for precision refinement. This approach balances computational efficiency with accuracy—filtering all 38,979 PRs using LLMs would be prohibitively expensive, so we first reduce the dataset with regex parsers before applying sophisticated LLM-based filtering.

We implemented a custom regex-based parser for TypeScript type constructs by consulting the official language documentation~\cite{typescript_handbook_2025}. Each parser examines three aspects of a PR:
\begin{enumerate}[leftmargin=*]
    \item \textbf{PR Title and Description:} We search for type-related keywords such as ``type'', ``generic'', ``nullable'', and ``\texttt{any}''.
    \item \textbf{Code Patches:} We analyze git diffs to detect type annotations, type declarations, and other type-related syntax.
    \item \textbf{File Extensions:} We verify that the PR modifies valid TypeScript (\texttt{.ts}, \texttt{.tsx}) files, excluding declaration files (\texttt{.d.ts}).
\end{enumerate}

Formally, let \(p\) denote a PR. We define \(\text{TS}_{File}(p)\) as PR that modifies at least one valid TypeScript file, \(\text{TS}_{Diff}(p)\) as the PR that contains type-related code modifications in git diff patches, and \(\text{FP}(p)\) as the PR that is a false positive. To mitigate false positives, we implemented language-specific exclusion rules. For TypeScript, we filter out HTML attributes (e.g., \texttt{type="button"}), MIME types, and logging statements. Hence, a PR \(p\) is finally classified as type-related if:
\[
\text{PR}_{Type}(p) = \text{TS}_{File}(p) \;\wedge\; \big( \text{TS}_{Diff}(p) \;\vee\; \text{TS}_{Cue}(p) \big) \;\wedge\; \neg \text{FP}(p)
\]

Many PRs, particularly the Agentic PRs, contain minimal or ambiguous descriptions. For example, a PR titled ``Fix bugs'' might involve significant type changes without explicitly mentioning types. Conversely, some PRs mention ``type'' but do not involve type system changes. To handle these cases, we also developed a multi-agent LLM-based pipeline after regex filtering. 

This pipeline consists of two agents: The classifier agent and the validator agent. The classifier agent receives the PR title, description, commit messages, code patches, and statistics (lines changed, files modified). We provide a detailed prompt defining type-related PRs with examples (adding type annotations, converting from \texttt{any} to specific types, fixing type errors, refactoring interfaces, etc.). The agent outputs a JSON response containing its boolean decision, a confidence score, as well as a detailed explanation of the decision with specific evidence. The validator agent then reviews the classifier agent's decision and checks for false positives, false negatives, and whether the confidence level is appropriate. A PR is included in our final dataset only if the validator agent approves the classifier. For both agents, we use OpenAI's \texttt{gpt-4o} model. Table~\ref{tab:dataset_size} shows the number of PRs at each filtering stage.

\begin{table}[h]
\caption{Number of PRs at each filtering stage}
\Description{A table summarizing the number of Pull Requests (PRs) for TS-AI Agent and TS-Human categories at three stages: Original Dataset, After Regex Parser, and After Multi-agent LLM. The counts decrease significantly at each stage.}
\centering
\renewcommand{\arraystretch}{0.9}
\begin{tabular*}{\columnwidth}{l@{\extracolsep{\fill}}lr}
\toprule
\textbf{Stage} & \textbf{Category} & \textbf{Count} \\
\midrule
\multirow{2}{*}{Original Dataset}
  & TS–AI Agent & 33,596 \\
  & TS–Human    & 5,080 \\
\midrule
\multirow{2}{*}{After Regex Parser}
  & TS–AI Agent & 1,083 \\
  & TS–Human    & 655 \\
\midrule
\multirow{2}{*}{After Multi-agent LLM}
  & TS–AI Agent & \textbf{545} \\
  & TS–Human    & \textbf{269} \\
\bottomrule
\end{tabular*}
\label{tab:dataset_size}
\end{table}

We evaluated each pipeline's performance by manually annotating 50 random TypeScript PRs. Two researchers with experience in TypeScript performed the annotation. We prioritized high recall over accuracy for the regex parser because our goal was to capture as many type-related PRs as possible, even if some false positives were included. The lower accuracy would be addressed in the LLM filtering stage. Our parser achieved a recall score of 0.85 and an accuracy score of 0.74. Due to the low accuracy of the first stage, we adopted the second multi-agent LLM stage, which achieved perfect accuracy and precision with current data and prompt settings after regex filtering. 

\section{Results}
\begin{table}[!t]
\caption{Mean number of unique advanced TypeScript type features used per PR.}
\Description{A table comparing the mean number of unique advanced TypeScript type features used per PR. It shows that AI agents (Claude\_Code, Cursor, Devin, Copilot, and OpenAI\_Codex) have higher mean usage, ranging from 5.50 to 6.74, whereas the mean for Human developers is significantly lower at 2.66.}
\centering
\renewcommand{\arraystretch}{0.9}
\begin{tabular*}{\columnwidth}{l@{\extracolsep{\fill}}r}
\toprule
\textbf{Developer / Agent} & \textbf{Mean Features} \\
\midrule
Claude\_Code     & 6.74 \\
Cursor           & 5.96 \\
Devin            & 5.79 \\
Copilot          & 5.57 \\
OpenAI\_Codex    & 5.50 \\
\midrule
Human            & 2.66 \\
\bottomrule
\end{tabular*}
\label{tab:feature_diversity}
\end{table}

\begin{figure*}[t]
    \centering
    \begin{minipage}[t]{0.49\textwidth}
      \centering
      \includegraphics[width=\linewidth]{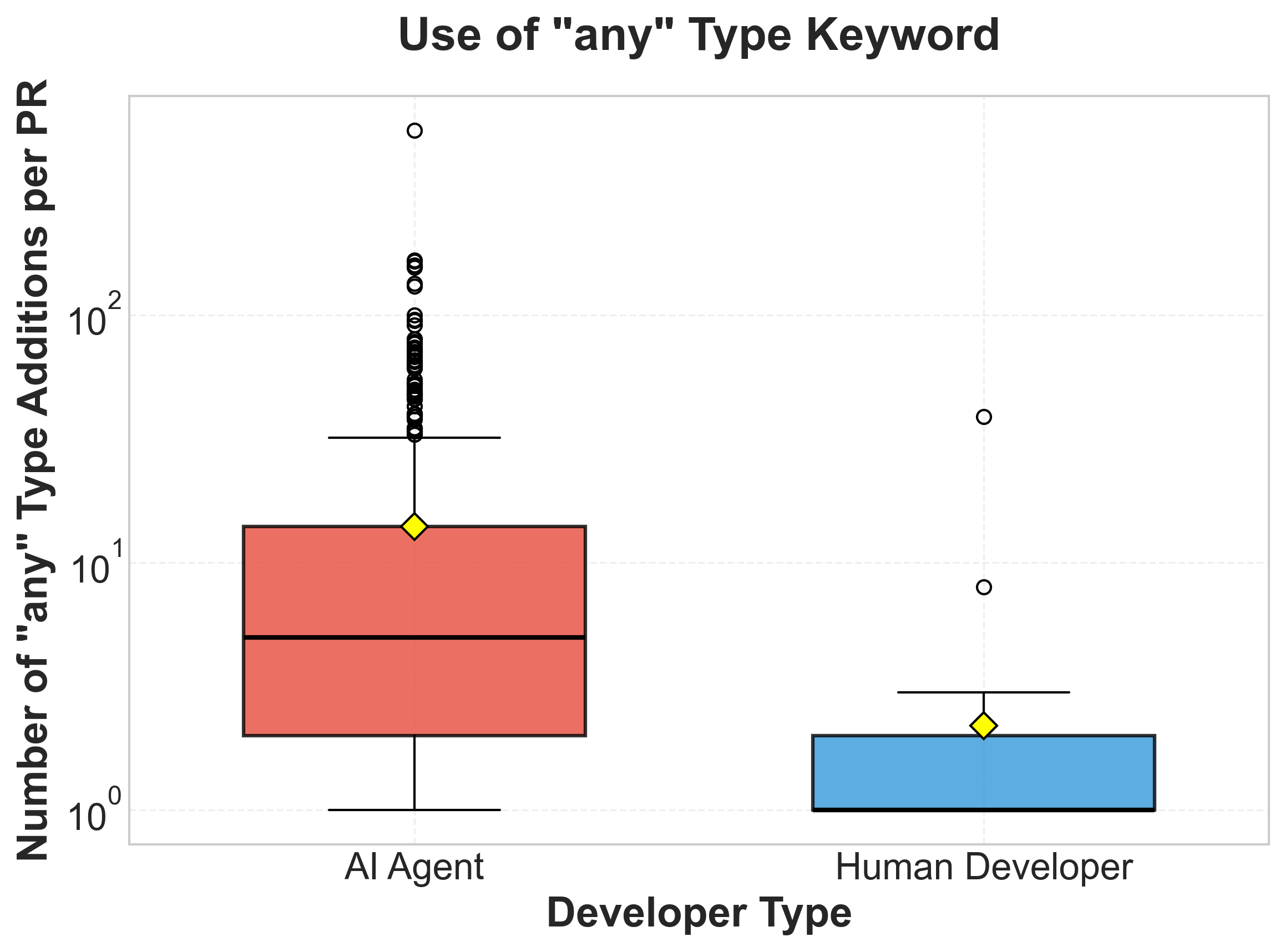}
      \caption{Comparison of AI and human type usage.}
      \Description{A figure comparing the number of 'any' type additions per PR between Agentic PRs and Human PRs. The AI Agents tend to add more 'any' type keywords compared to human developers significantly.}
      \label{fig:type_comparison}
    \end{minipage}
    \hfill
    \begin{minipage}[t]{0.49\textwidth}
      \centering
    \includegraphics[width=\columnwidth]{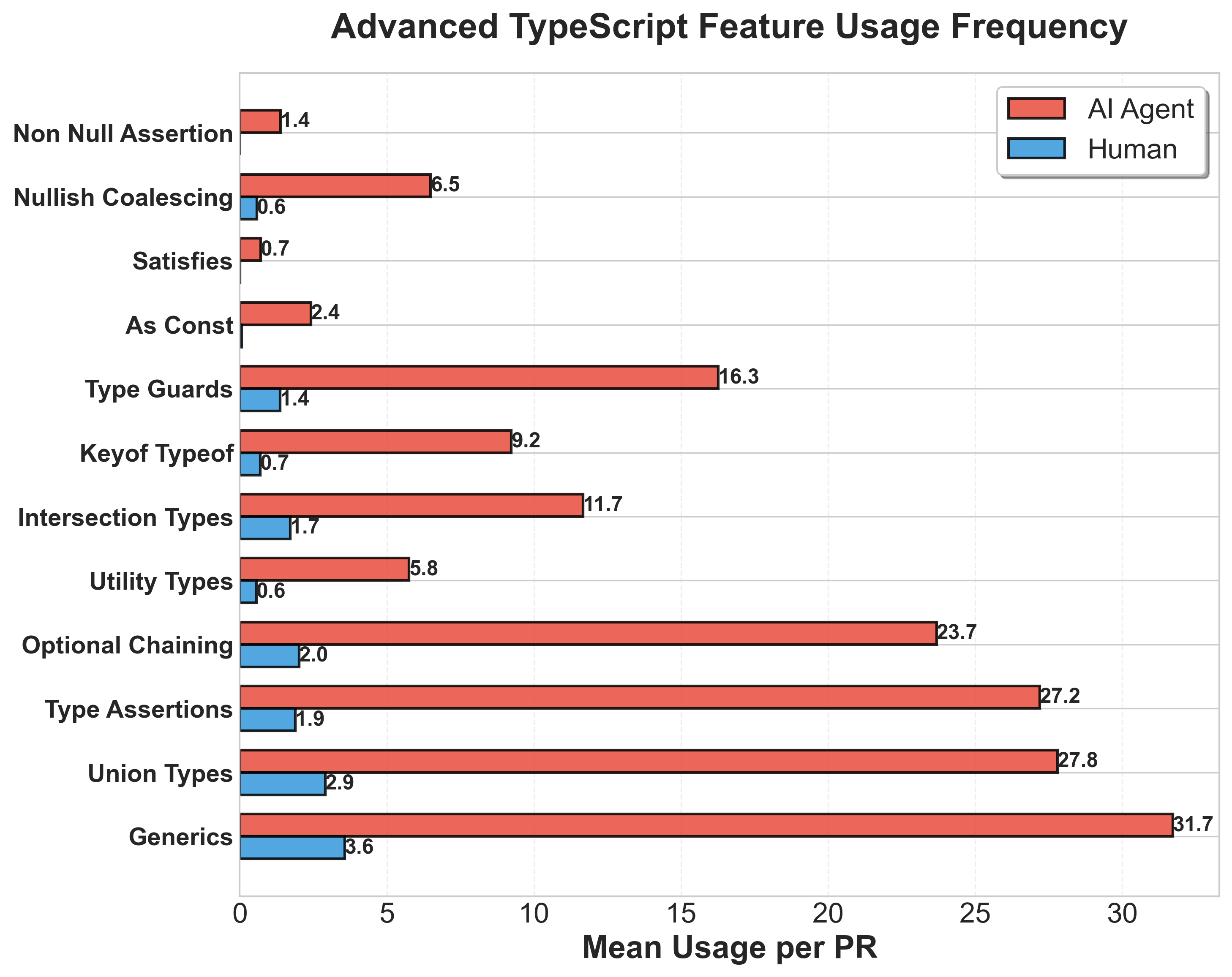}
    \caption{Advanced TypeScript type construct usage by agent.}
    \Description{A figure showing the advanced TypeScript type constructs usage with a comparison between AI agents and human developers. AI agents use type constructs more often, even including anti-patterns such as non-null assertions and type assertions.}
    \label{fig:feature_usage_by_agent}
    \end{minipage}
\end{figure*}

\begin{figure}[t]
    \centering
    \includegraphics[width=\columnwidth]
    {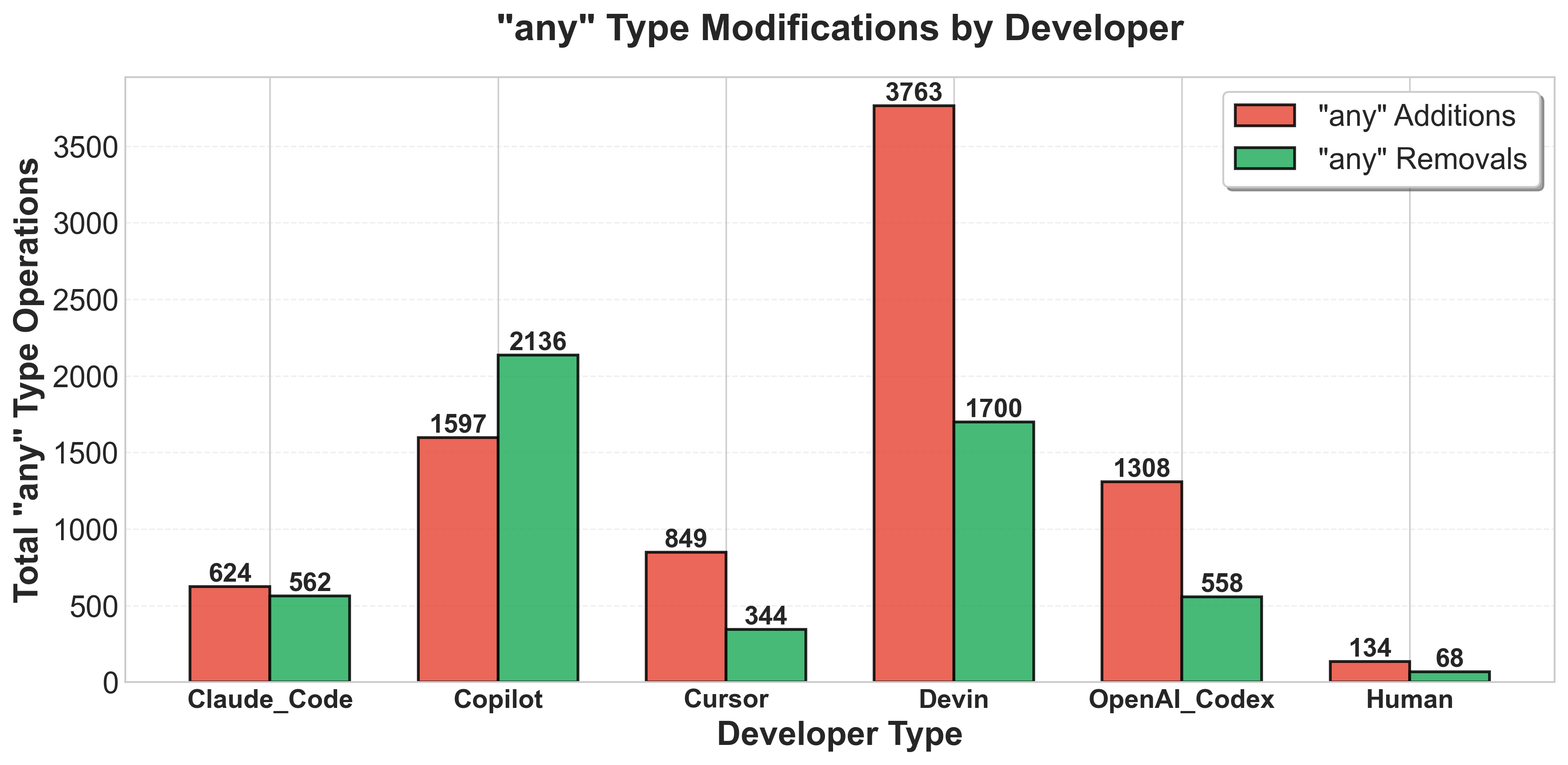}
      \caption{'any' type additions and removals by agent.}
      \Description{A figure showing the number of 'any' type keyword additions and removals by agent. Devin shows the biggest number of 'any' additions, while Copilot records the highest number of 'any' removals, even more than additions.}
      \label{fig:type_comparison_by_agent}
\end{figure}

{\textbf{\textit{RQ1:} Do AI agents actually introduce type-related issues?
If so, how frequently do they introduce them?}} 

As discussed in the background section, if either AI agents or human developers introduce the use of \texttt{any} type keywords instead of a concrete type within TypeScript, respectively, then this indicates potential runtime issues. We compared the usage behavior of these type-related keywords from humans and AI agents to see the pattern. As shown in Figure~1, AI agents are 9 times more likely to introduce the \texttt{any} type into the TypeScript codebases compared to human developers, with means of the `any addition' distributions across PRs at 2.16 and 0.24, respectively. Mann-Whitney U test ($p \approx 2.33 \times 10^{-7}$) demonstrated the statistical significance of this result (Cohen's $d = 0.32$).

Figure~3 shows additions and removals of \texttt{any} type keywords per agent in the TypeScript codebases. Suppose an AI agent introduces more \texttt{any} additions and very few \texttt{any} removals, then it induces the most amount of type-related runtime issues by keeping a lot of \texttt{any} keywords in the codebases. In Figure~3, even though Devin has the highest number of \texttt{any} keyword additions due to its presence in more PRs compared to other agents, the ratio of 
\(\texttt{any\_additions} / \texttt{any\_removals}\) is the worst in the case of Cursor (2.46). OpenAI Codex follows next with the ratio of 2.34. GitHub Copilot shows a tendency to remove \texttt{any} types overall, whereas other AI agents also remove \texttt{any} types but add them more.

Based on our findings and analysis, we conclude the answer to RQ1 as: AI agents tend to introduce \texttt{any} type keyword rather than reduce them in a statistically significant manner in TypeScript. More broadly, AI agents show different behaviors across the models, but are still highly prone to introducing \texttt{any} type keywords.

\noindent{\textbf{\textit{RQ2:} How do AI agents and human developers differ in their use of type-related features?}} 

Table~2 shows that AI agents employ advanced type-related features more often compared to human developers. This observation is clear from the higher mean usage of advanced TypeScript type-related features, including type-related patterns and anti-patterns per PR by the AI agents. Figure~2 shows all the specific advanced type-related features that we analyzed for TypeScript. Clearly, for all the type constructs, the Agentic PRs are more prone to introducing advanced type-related features compared to humans. According to the Mann-Whitney U test, this finding is quite statistically significant (with a p-value of less than $5.50 \times 10^{-5}$, Cohen's $d = 1.45$). Figure~2 presents an additional interesting insight that Agentic PRs are also more likely to introduce overly type-related anti-patterns, such as non-null assertions and type assertions. These patterns allow the code to bypass crucial checks during compile-time, and thus, they must be used with great caution. Even then, according to our results, the usage frequency of these type-related anti-patterns is much larger within Agentic PRs compared to Human PRs for TypeScript. This again indicates that AI agents tend to use a broader set of advanced type constructs, even when those constructs may not be necessary.

\noindent\textbf{\textit{RQ3: }How do Agentic PRs compare to human developers in acceptance rate?}

So far, we have observed that Agentic PRs introduce substantially more type-related issues than Human PRs. Only under this observation, one might expect Agentic PRs to exhibit lower merge rates. According to our result, Agentic PR's rejection rate is 42.6\% while Human PR's rate is only 2.6\%. However, surprisingly, the result also shows that AI agents achieve significantly higher acceptance rates than human developers. With a statistically significant difference (\(p < 0.0001\), \(\chi^2 = 27.52\), Cramér's $V = 0.32$), the acceptance rate of Agentic PRs is 45.8\%, while that of Human PRs is only 25.3\% in type-specific PRs. From the analysis of results in RQ2, we find that AI agents consistently introduce type-related anti-patterns and overly complex type constructs. We know that this acceptance-rate gap should be interpreted with caution, as Agentic PRs and Human PRs may differ systematically in the types of tasks they address. Although the higher acceptance rate of Agentic PRs does not necessarily imply that reviewers overlook these issues, our results suggest that type-related problems introduced by AI-generated TypeScript code may be more easily accepted in certain contexts, potentially accumulating as technical debt over time. If AI agents are heavily relied upon during development, such debt may increase the long-term maintenance burden for human developers.

\begin{table}[t]
\caption{Acceptance rates of TypeScript by AI agents.}
\centering
\renewcommand{\arraystretch}{0.87}
\begin{tabular*}{\columnwidth}{l@{\extracolsep{\fill}}r}
\toprule
\textbf{Developer/Agent} & \textbf{Acceptance Rate (\%)} \\
\midrule
OpenAI\_Codex & 60.3 \\
Cursor & 50.6 \\
Claude\_Code & 50.0 \\
Devin & 42.4 \\
Copilot & 33.0 \\
Human & \textbf{25.3} \\
\bottomrule
\end{tabular*}
\label{tab:acceptance_rates}
\end{table}

\section{Threats to Validity}
We acknowledge that our data filtering framework relies on custom regex-based parsers to identify type-related PRs. While these parsers cover a comprehensive set of TypeScript type features, regex patterns inherently lack the semantic awareness of a full abstract syntax tree (AST). Our parsers may miss unconventional syntax or deeply nested generic structures (false negatives). Although we mitigated false positives through a subsequent multi-agent LLM validation stage, the initial reliance on regex means that our analysis is strictly bound to the predefined patterns and may under-report the usage of novel or complex type features not captured by our expressions. Moreover, we acknowledge that probabilistic biases inherent to LLMs may influence the dataset refinement, even though the current accuracy of our multi-agent LLM filtering system is 1.0 with our prompt. It is important that the influence of user-specified type safety requirements on agent performance remains outside our study's scope.

\section{Ethical Implications}
This study analyzes publicly available PR data from the AIDev dataset and does not involve direct interaction with human participants. No personally identifiable information (PII) is collected or reported, and all analyses are performed at an aggregate level without profiling individual developers. Consistent with ACM policy, our research isolates systematic differences between AI-generated and human-authored code.

\section{Conclusion}
This study presents the first empirical investigation into how AI agents navigate hybrid type systems in TypeScript. Our analysis reveals that Agentic PRs introduce type-related anti-patterns and escape hatches like \texttt{any} type keyword significantly more often than human developers. AI agents tend to actively use not only type-related patterns but also anti-patterns that allow bypassing type safety checks, leading to potential runtime errors. In particular, AI agents are approximately 9$\times$ more likely to introduce the \texttt{any} type compared to human developers (mean additions per PR: 2.16 vs. 0.24). Despite these issues, Agentic TypeScript PRs exhibit substantially higher acceptance rates than Human PRs. In our dataset, Agentic PRs achieved an acceptance rate of 45.8\%, whereas Human PRs were accepted at a rate of only 25.3\%. This contradiction suggests that repositories might be unintentionally accumulating technical debt. Our findings indicate that current AI agents prioritize compilation success over long-term type soundness and maintainability. We recommend that teams adopting AI-assisted development enforce stricter type-focused review guidelines to detect risky type constructs early. 

\begin{acks}
Our work was funded by the Natural Sciences and Engineering Research Council of Canada (NSERC).
\end{acks}

\bibliographystyle{ACM-Reference-Format}
\bibliography{references}

\appendix

\end{document}